\documentclass[prb,showpacs,twocolumn,preprintnumbers,amsmath,amssymb,floatfix]{revtex4}
\usepackage[dvips]{graphicx}
\usepackage[latin1]{inputenc}
\begin{document}
\draft

\newcommand{\lpc} {Li$_{0.5}$MnPc}
\newcommand{\lisi} {Li$_2$VOSiO$_4$}
\newcommand{\etal} {{\it et al.} }
\newcommand{\ie} {{\it i.e.} }
\newcommand{\aucr}{CeCu$_{5.9}$Au$_{0.1}$ }
\newcommand{\auaf}{CeCu$_{5.2}$Au$_{0.8}$ }
\newcommand{\aux}{CeCu$_{6-x}$Au$_{x}$ }
\newcommand{\ip}{${\cal A}^2$ }

\hyphenation{a-long}

\title{Low-energy excitations in electron-doped metal phthalocyanine \\ from NMR in Li$_{0.5}$MnPc }

\author{M.Filibian, and P. Carretta}

\affiliation{Department of Physics ``A.Volta", University of
Pavia, Via Bassi 6, I-27100, Pavia (Italy)}

\author{T. Miyake, Y. Taguchi, and Y. Iwasa}

\affiliation{Institute for Materials Research, Tohoku University, Sendai 980-8577 (Japan)}

\widetext

\begin{abstract}

$^7$Li and $^1$H NMR and magnetization measurements in \lpc\, (Pc$\equiv$C$_{32}$H$_{16}$N$_8$), recently proposed
as a strongly correlated metal, are presented. Two different low-frequency dynamics are evidenced. The first one,
probed by $^1$H nuclei gives rise to a slowly relaxing magnetization at low temperature and is associated with the
freezing of MnPc $S=3/2$ spins. This dynamic is similar to the one observed in pristine $\beta$-MnPc and
originates from Li depleted chain segments. The second one, evidenced by $^7$Li spin-lattice relaxation rate, is
associated with the hopping of the electrons along Li-rich chains.  The characteristic correlation times for the
two dynamics are derived and the role of disorder is briefly discussed.

\end{abstract}

 \maketitle

\narrowtext

\section{Introduction}

Metal phthalocyanines (hereafter MPc) have attracted a lot of interest in the last decades owing to their
technological applicabilities as dyes, gas sensors or in electro-optical devices \cite{tec}. A renewed interest on
these systems has emerged after the observation that thin films of MPc show a marked increase in the electrical
conductivity once they are doped with alkali ions \cite{Morpurgo}. In view of the similarities that alkali-doped
MPc (A$_x$MPc) share with fullerides, Tosatti et al. \onlinecite{Tosatti} have analyzed, within a model
successfully applied to other strongly correlated electron systems \cite{Capone}, the possibility that
superconductivity could occur also in A$_x$MPc.  It was found that strongly correlated superconductivity could
develop also in A$_x$MPc for $x\simeq 2$, with a magnitude and a symmetry of the order parameter which would
depend on the intensity of the local exchange (Hund coupling) and on Jahn-Teller coupling \cite{Tosatti}.
Accordingly a growing interest on these compounds has arisen in the last year. Nevertheless, the synthesis of bulk
A$_x$MPc is non-trivial and hitherto still at a preliminary stage. So far only $\beta$-Li$_x$MnPc powders have
been grown in a reproducible way, for $0\leq x\leq 4$ \cite{Yasu}, and the evolution of the lattice parameters and
of their basic magnetic properties with doping have been studied .

\begin{figure}[h!]
\vspace{6cm} \includegraphics{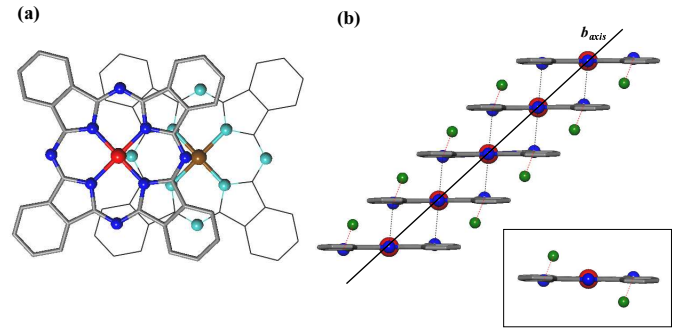} \caption{(a) Projection of
two neighboring MnPc molecules within the same stack along the normal direction to their planar rings in
Li$_2$MnPc. The Mn2+ ions are depicted as red (brown) spheres lying directly above (below) neighboring light blue
(dark blue) N atoms. (b) The slip-stacked MnPc assembly along the b axis in Li$_2$MnPc. Li$^+$ ions (depicted in
green) are disordered and reside exclusively in the intrastack spacing and bond strongly to pyrrole-bridging N
atoms (red dotted lines). The inset shows the Li$_2$MnPc building block of the 1D assemblies. }\label{figsquid}
\end{figure}

Li$_x$MnPc structure (Fig.1) is formed by chains along which MnPc molecules are stacked. From high-resolution
X-ray diffraction it was observed that Li$^+$ ions stay in between adjacent molecules piling up along the chain
and are tightly bound to pyrrole-bridging N atoms \cite{Yasu}. This one-dimensional structure is quite similar to
the one of other organic conductors which have been deeply investigated in the last twenty years and are still
subject of an intense research activity, the Beechgard salts \cite{Beechgard}. Magnetization measurements carried
out in Li$_x$MnPc, have revealed a progressive increase in the Curie constant with doping and a modification in
the magnitude and sign of the superexchange coupling, a neat demonstration that electrons are transferred to MPc
molecular orbitals. In order to study the modifications of the microscopic electronic properties of Li$_x$MnPc
with doping, one can conveniently use local probes as $^7$Li and $^1$H nuclei. In the following we present a
$^7$Li and $^1$H nuclear magnetic resonance (NMR) and magnetization study of Li$_{0.5}$MnPc. From nuclear
spin-lattice relaxation measurements two different dynamics were found. A low-frequency dynamics associated with
the progressive freezing of MnPc $S=3/2$ spins and a dynamics at much higher frequency of diffusive character. The
characteristic correlation times, the hyperfine couplings and the temperature and time evolution of the
macroscopic magnetization are estimated and discussed in the light of the possible evolution of the molecular
electronic and spin configuration. The experimental results are presented in the following sections. Section III
is dedicated to their analysis and discussion, while the final conclusions are presented in Sect. IV.

\section{Technical Aspects and Experimental results}

\begin{figure}[h!]
\vspace{10cm} \includegraphics{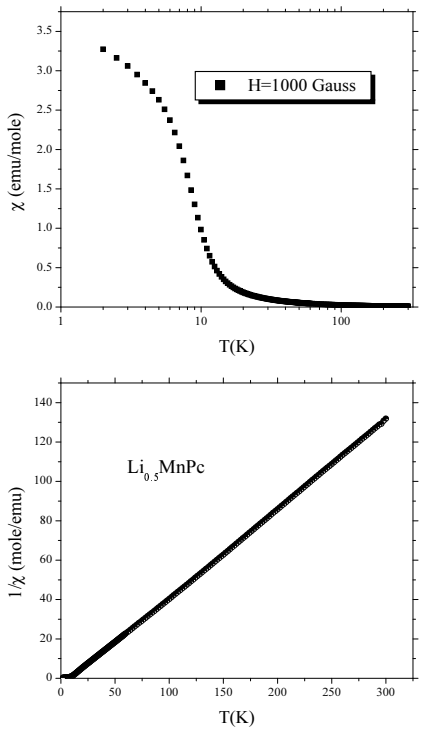} \caption{(top) Temperature
dependence of the susceptibility, defined as $\chi=M/H$, in \lpc for $H=1$ kGauss. (bottom) Temperature dependence
of the inverse of the susceptibility. One can notice that for $T\geq 15$ K Curie-Weiss law is
obeyed.}\label{figsquid}
\end{figure}

\begin{figure}[h!]
\vspace{6cm} \includegraphics{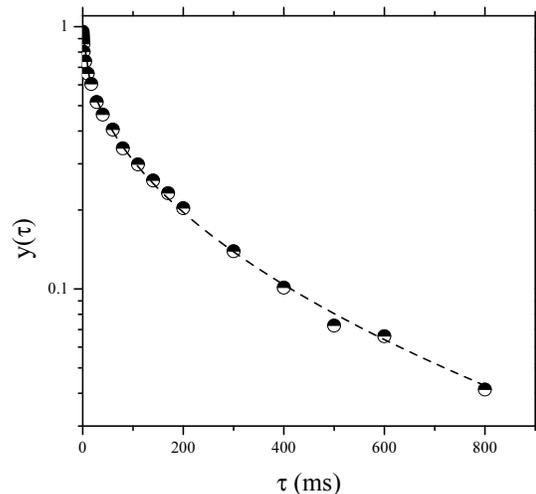} \caption{Recovery law for
$^7$Li nuclear magnetization after a saturating pulse sequence. The dashed line is the best fit according to
$y(\tau)=exp(-(\tau/T_1)^{\beta})$}\label{Rec1H}
\end{figure}

As-purchased MnPc powder was purified by vacuum sublimation prior to the intercalation procedure. Li-intercalation
was carried out by using liquid-phase process in an Ar-filled glove box. Details of the sample preparation
procedures are described in Ref.~\onlinecite{Yasu}. The powder samples were then sealed in a quartz tube  to avoid
oxidation.

Magnetization measurements were performed using a Quantum Design XPMS-XL7 SQUID magnetometer. At high
temperatures, above $15$ K, the magnetization $M$ was found to increase linearly with the field intensity $H$ and
hence the susceptibility can be defined from $\chi= M/H$. One observes (Fig.2) a high temperature Curie-Weiss
behaviour
\begin{equation}
\chi= \frac{C}{T-\Theta} + \chi_{o} ,
\end{equation}
with $C= 2.27$ emu.K/mole Curie constant and $\Theta=7.8 \pm 0.2$ K, indicating a dominant ferromagnetic coupling.
$\chi_{o}$ is the sum of Van-Vleck and diamagnetic contributions, which are assumed weakly temperature dependent
\cite{Barraclough}. Below about 10 K a clear departure from Curie-Weiss law is observed. The magnetization shows
an upturn, is no longer linear in the field and is observed to slowly relax in time, a behaviour suggesting a
freezing of the molecular spins. This behaviour is reminiscent of the one observed in pure $\beta$-MnPc
\cite{Yasu}.

\begin{figure}[h!]
\vspace{6cm} \includegraphics{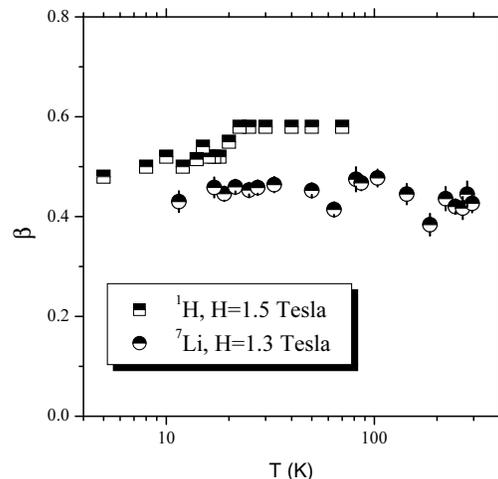} \caption{Temperature
dependence of the stretched exponential exponent $\beta$ for $^7$Li and for the fast component of $^1$H recovery
laws.}\label{betas}
\end{figure}

NMR measurements were performed by using standard radiofrequency (RF) pulse sequences. $^7$Li and $^1$H powder
spectra were obtained from the Fourier transform of half of the echo signal after a $\pi/2-\tau -\pi/2$ pulse
sequence. The spectra were observed to be Gaussian with a linewidth increasing upon cooling, as the macroscopic
susceptibility. In $^7$Li ($I=3/2$) spectra there was no clear evidence of the satellite transitions, which are
much less intense than the central one. Moreover, we observed that the length of the $\pi/2$ pulse was about half
of that derived for $^7$Li in an acqueous solution of LiCl, where all transitions are irradiated. This indicates
that practically just the central $m_I=1/2 \leftrightarrow -1/2$ transition of $^7$Li was irradiated. The echo
intensity $E(2\tau)$ was observed to decrease upon increasing $\tau$ following an almost Gaussian law, with a
characteristic decay time for $^7$Li around $T_2^G\simeq 165 \mu s$, while $\simeq 45 \mu s$ for $^1$H, around 100
K. Finally, the relative intensity of $^7$Li and $^1$H signals for $\tau\rightarrow 0$ was observed to be
consistent with a Li content of $0.50 \pm 0.01$ per formula unit, in excellent agreement with the expected nominal
doping.

\begin{figure}[h!]
\vspace{6cm} \includegraphics{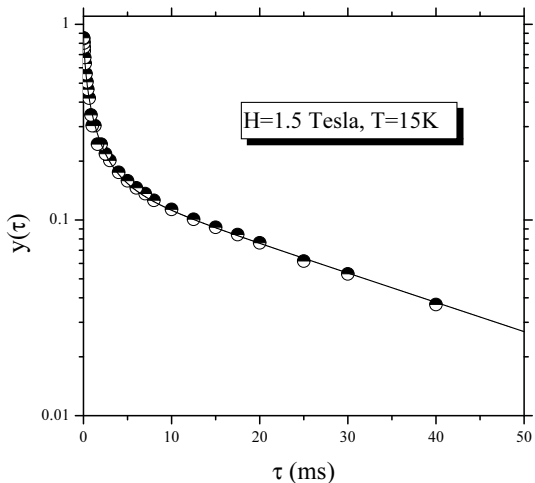} \caption{Recovery law for
$^1$H nuclear magnetization after a saturating pulse sequence. The solid line is the best fit according to
Eq.2}\label{Rec1H}
\end{figure}

Nuclear spin-lattice relaxation rate $1/T_1$ was estimated from the recovery of nuclear magnetization $m(\tau)$
after a saturating RF pulse sequence. The recovery law of $^7$Li was observed to be a stretched exponential (Fig.
3), namely $y(\tau)\equiv 1- m(\tau)/m(\tau\rightarrow \infty)= exp(-(\tau/T_1)^{\beta})$, with $\beta\simeq 0.45$
over all the temperature range (Fig. 4). A stretched exponential recovery indicates the presence of disorder at
the microscopic level. Also $^1$H recovery law was essentially of stretched exponential character, however, for
large delays $\tau$ a clear departure from a simple stretched exponential recovery was noticed. The recovery could
be nicely fit according to (Fig. 5)
\begin{equation}
y(\tau)= A e^{-(\frac{\tau}{T_1^s})^{\beta}}+ (1-A) e^{-(\frac{\tau}{T_1^l})}
\end{equation}
with $A\simeq 0.8$ and $\beta\simeq 0.5$ over the explored temperature range (Fig. 4). The temperature dependence
of $^7$Li and $^1$H relaxation rates derived from the aforementioned recovery laws are shown in Figs. 6 and 8. At
temperatures above 35 K $^1H$ relaxation rate $1/T_1^s$ shows a low-frequency divergence with $1/T_1^s\propto
1/\sqrt{H}$ (Fig. 7), a behaviour which is typical of one-dimensional spin systems.

\begin{figure}[h!]
\vspace{6.5cm} \includegraphics{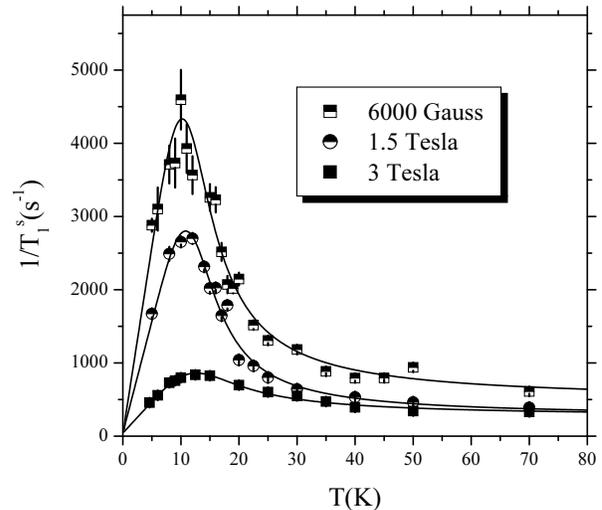} \caption{Temperature
dependence of  $^1$H $1/T_1^s$ at different magnetic fields. The solid lines are the best fit according to Eq.8,
for $<E_A>\simeq \Delta= 25$ K.}\label{T1H}
\end{figure}

\begin{figure}[h!]
\vspace{6cm} \includegraphics{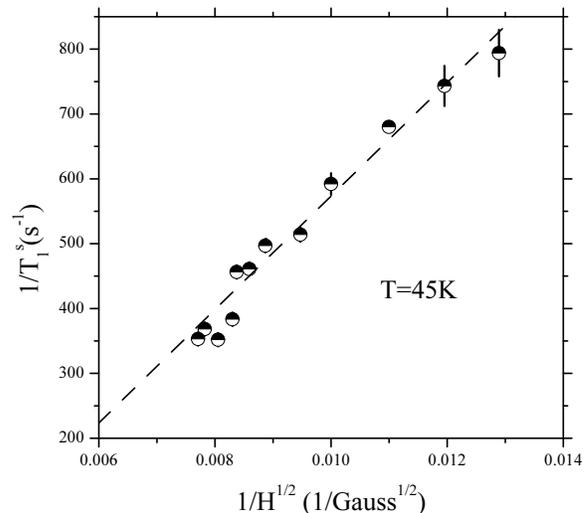} \caption{$^1$H nuclear
spin-lattice relaxation rate $1/T_1^s$, at $T=45$ K, reported as a function of $1/\sqrt{H}$ in order to evidence
the diffusive nature of the spin correlations. The dashed line is the best fit according to Eq. 6 in the
text.}\label{T1HvsH}
\end{figure}

\section{Discussion}

\begin{figure}[h!]
\vspace{6cm} \includegraphics{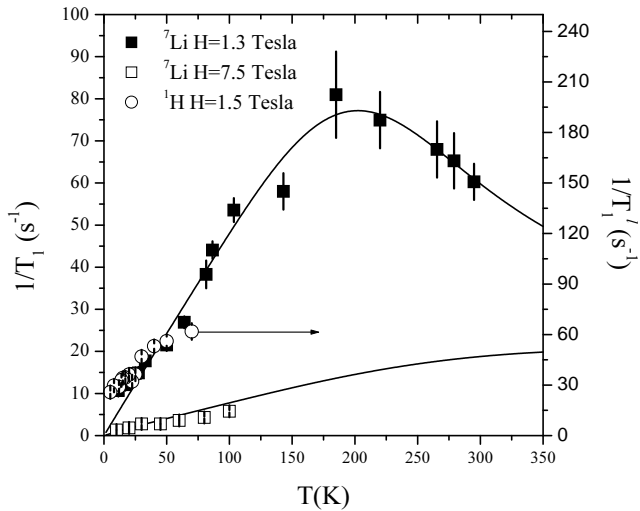} \caption{Temperature
dependence of  $^7$Li $1/T_1$ (squares) and of $^1$H $1/T_1^l$ (circles and right vertical scale), the relaxation
rate of the slowly relaxing component of $^1$H magnetization (see Fig.5). The solid lines are the best fit
according to Eq.8, for $<E_g>\simeq \Delta= 410$ K.}\label{T1H}
\end{figure}

The temperature dependence of the magnetic susceptibility is characteristic of ferromagnetically correlated spin
chains, as observed for pristine $\beta$-MnPc. The fit of the susceptibility according to Curie-Weiss law yields
$\Theta= 7.5$ K, corresponding to an exchange coupling constant $J=\Theta/ [2z\sqrt{S(S+1)/3}]\simeq 1.7$ K, for
$S=3/2$ and taking $z=2$ for the number of nearest neighbours. This value of $\Theta$ is lower than the one of
pure $\beta$-MnPc, whereas the Curie constant is larger in \lpc , in agreement with the results reported by
Taguchi et al. \onlinecite{Yasu} for $0\leq x\leq 4$ who observed that for $x\rightarrow 2$ Curie constant tends
to the one expected for $S=5/2$ with $g=2$, while $\Theta$ becomes negative, i.e. the chain becomes
antiferromagnetically correlated.

$\beta$-MnPc in its ground-state has a completely filled $a_{1u}$ level, a half-filled $a_{1g}$ level and two
degenerate half-filled 1$e_g$ levels \cite{electro,Barraclough}. In order to attain an $S=5/2$ molecular spin
state upon increasing Li content it is necessary to start filling the upper energy levels, as the two-fold
degenerate $2e_g(\pi)$ LUMO and the $b_1(d_{x^2-y^2})$ states. A scenario which differs from the one of MPc
containing transition metal ions with more d-electrons than Mn$^{2+}$. In fact, in FePc and CoPc Hund rule is not
obeyed and the other electrons occupy the 1$e_g$ levels, so that one has an $S=1$ and an $S=1/2$ configuration,
respectively. Taking into account the crystal-field modifications induced by Li$^+$, the $S=5/2$ state could in
principle result from two different configurations. As suggested in Ref.\onlinecite{Yasu}, Li$^+$ gives rise to a
more isotropic crystal-field and the energy of the LUMO and $b_1(d_{x^2-y^2})$ can be lowered, so that an electron
can be promoted from the $a_{1u}$ to the $b_1(d_{x^2-y^2})$ level, in accordance to Hund's rule. Then the $S=5/2$
state would result from the occupancy of all $d$-character orbitals. The electrons injected by Li doping, which
fill $2e_g(\pi)$ LUMO, would not contribute to the total molecular spin once they form a singlet state, a
situation similar to the one observed in the fullerides \cite{Forro}. Singlet formation can occur if Jahn-Teller
coupling for the $2e_g(\pi)$ state with the $B_{1g}$ and $B_{2g}$ $\beta$-MnPc vibrational modes is larger than
its Hund coupling \cite{Tosatti}.

On the other hand, an $S=5/2$ state is also consistent with a triplet state for the electrons on the LUMO,
provided that $b_1(d_{x^2-y^2})$ energy is not significantly lowered by Li doping and this latter level remains
empty. The comprehension of which one of the two configurations is actually taking place is important, as it will
allow to know which one of the two couplings, Jahn-Teller or Hund's, is dominant for the electrons on the LUMO and
then, in case strongly correlated superconductivity would develop, the symmetry of the superconducting order
parameter. Nevertheless, from our measurements alone it is not possible to discern among the two scenarios.
Crystal-field calculations and microscopic techniques using Mn$^{2+}$ ions as probes (e.g. EPR), would allow to
clarify this aspect. It is noticed that the filling of the LUMO, which overlaps with the $a_{1g}$ orbitals of
adjacent molecules along the chain, can induce antiferromagnetic correlations \cite{Evangelisti}, which would
explain the change of sign of Curie-Weiss temperature with increasing doping.

Now, the configuration with two Li$^+$ ions close to an MnPc molecule gives rise to a crystal-field which decrease
the energy of the $b_1$ states and also of the LUMO. Accordingly, one should expect that the energy of $S=5/2$
state is possibly lower than the one of $S=2$ and $S=3$ states, associated with one unpaired electron on the LUMO.
Then, it is likely that the progressive increase in the Curie constant, upon increasing Li content from $x=0$ to
2, actually originates from the contribution of $S=3/2$ and of $S=5/2$ chain segments, which are characterized by
a more stable spin configuration. Although from magnetization data alone one cannot say if both spin
configurations coexist at the microscopic level, we remark that Li$_x$MnPc susceptibility data can be fit with
$\chi= (x/2)\chi_{Li_2} + [(2-x)/2]\chi_{Li_0}$, for $T\gg 10$ K, where the $\chi_{Li_{x=0,2}}$ susceptibility are
the ones derived by Taguchi et al. \cite{Yasu} for $x=0$ and $x=2$. As will be pointed out later on, this
interpretation is corroborated by the observation of two different dynamics in nuclear spin-lattice relaxation
measurements.

At low temperature one observes a deviation from Curie-Weiss law which, however, is not associated with a
three-dimensional long-range order. In fact, one observes a slowly relaxing magnetization at low temperature which
suggests a progressive freezing of the molecular spins, a scenario observed in other spin chains characterized by
a sizeable magnetic anisotropy \cite{Bogani}. At a temperature of $3$ K the magnetization was observed to recover
its equilibrium value, after increasing the field from 0 to $1$ Tesla, in a characteristic time $\tau_M\simeq
1050$ s. This macroscopic relaxation has to be associated with the long-wavelength $q\rightarrow 0$ modes. One can
compare this result with the average relaxation time measured by means of AC-susceptibility by Taguchi et al.
\onlinecite{Yasu}. From the peak in the imaginary part of the AC-susceptibility measured at different frequencies
one can roughly estimate an average correlation time $\tau_{AC}\simeq 10^{-10} exp(90/T)$ s, corresponding to
$\tau_{AC}\simeq 1070$ s at 3 K, in rather good agreement with the value estimated from the relaxation of the
magnetization.

\begin{figure}[h!]
\vspace{6cm} \includegraphics{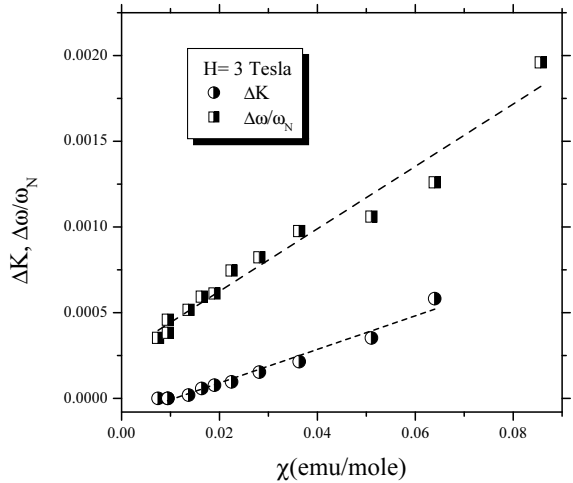} \caption{$^1$H paramagnetic
shift $\Delta$K (see Eq.4) and normalized line broadening $\Delta\omega/\omega_N$ reported as a function of the
magnetic susceptibility measured with the SQUID magnetometer, with the temperature as an implicit parameter. The
dashed lines evidence the linear dependence of the line broadening and of the shift on the susceptibility (see
Eq.4). }\label{Shift1H}
\end{figure}

Now we turn to the analysis of the low-energy excitations in the light of nuclear spin-lattice relaxation rate
$1/T_1$ measurements. In the presence of a magnetic relaxation mechanism
\begin{equation}
\frac{1}{T_1}= \frac{\gamma^2}{2}\int_{-\infty}^{+\infty}\, e^{i\omega_Nt}<h_+(t)h_-(0)>dt \,\,\, ,
\end{equation}
where $h_{\pm}$ are the transverse components of the hyperfine field at the nucleus, $\omega_N$ nuclear resonance
frequency and $\gamma$ the nuclear gyromagnetic ratio. For localized spins one has that $\vec h(t)= \sum_i
\tilde{A}_i \vec S_i(t)$, with $\tilde{A}_i$ the hyperfine coupling between the $i$-th electron spin and the
nucleus. In a single crystal the magnitude of the hyperfine coupling can be estimated from the shift of the
resonance frequency $\omega_N$ with respect to the nuclear Larmor frequency $\omega_o$ in the applied magnetic
field. In fact, one can write
\begin{equation}
\frac{\omega_N-\omega_o}{\omega_o}\equiv \Delta \tilde{K} = \frac{\tilde{A}_T\tilde{\chi}}{g\mu_BN_A}
\end{equation}
where the macroscopic susceptibility $\tilde{\chi}$, the hyperfine coupling $\tilde{A}_T= \sum_i \tilde{A}_i $ and
the shift $\Delta \tilde{K}$ are all tensors, in principle. Therefore, in a powder one should observe an effective
broadening of the line rather than a shift. Indeed both for $^1$H and $^7$Li one observes that the line broadens
with increasing $\chi$ (Figs. 9 and 10). The line shape is Gaussian for both nuclei and no structure associated
with the form of the shift tensor can be discerned. Therefore, by plotting the line broadening $\Delta\omega$ vs.
$\chi$, with the temperature as an implicit parameter, one can estimate an effective coupling constant $A$. For
both nuclei one finds $A\simeq 100$ Gauss, which is of the order of magnitude of the dipolar hyperfine field
generated by the localized spins. It is noticed that $^1$H spectra shows also a slight shift on cooling (Fig. 9),
still proportional to $\chi$ and yielding an isotropic hyperfine coupling constant still around $100$ Gauss.

\begin{figure}[h!]
\vspace{6cm} \includegraphics{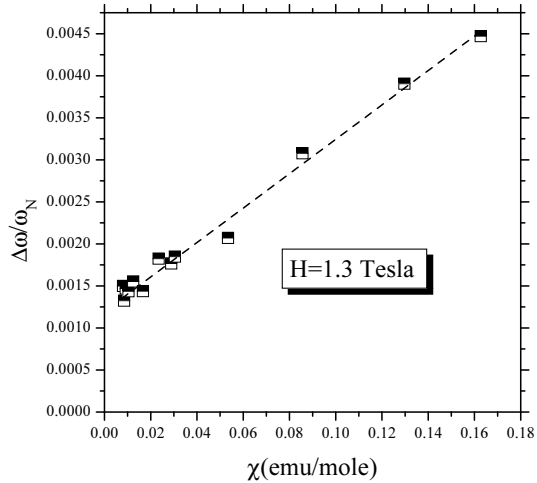} \caption{$^7$Li normalized
line broadening $\Delta\omega/\omega_N$ reported as a function of the magnetic susceptibility measured with the
SQUID magnetometer, with the temperature as an implicit parameter. }\label{Shift7Li}
\end{figure}

In the high temperature limit $T>>J$ the spins are uncorrelated and, in the absence of diffusive processes, one
would expect a temperature and field independent $1/T_1$, with \cite{Boucher}
\begin{equation}
\frac{1}{T_1}= \gamma^2 A^2 \frac{S(S+1)}{3} \frac{\sqrt{2\pi}}{\omega_H}
\end{equation}
where $\omega_H= (2zJ/k_B)\sqrt{S(S+1)/3}$ is the Heisenberg exchange frequency. Accordingly $^1$H $1/T_1$ is
estimated around $23$ s$^{-1}$ and $^7$Li $1/T_1$ around 3.5 s$^{-1}$, for $T\gg \Theta$. However, one notices
that for $T\gg 10$ K $^1$H $1/T_1^s$ is at least an order of magnitude larger than expected, moreover, it is
strongly field dependent (Figs. 6 and 7). The marked field dependence of $1/T_1^s$ for $T\gg \Theta$ can be
associated with the diffusive character of the spin correlations, which in one-dimension give rise to a
$1/\sqrt{H}$ low-frequency divergence in the spectral density \cite{Boucher}. In fact, by neglecting low-frequency
cutoffs, one can write
\begin{equation}
\frac{1}{T_1}= \gamma^2 A^2 \frac{S(S+1)}{3} \frac{1}{\sqrt{2D}} \frac{1}{\sqrt{\omega_N}} .
\end{equation}
By fitting the data in Fig.7 with the above equation one estimates a diffusion constant $D\simeq 1.2 \omega_H$, of
the right order of magnitude. It is noticed that upon cooling, for $T\rightarrow \Theta$, $1/T_1^s$ rapidly
increases and then shows a broad maximum around 10 K. The amplitude of the maximum is strongly field dependent and
is not associated with a three-dimensional ordering, but rather signals a low-frequency dynamics at frequencies of
the order of $\omega_N\ll \omega_H$. This maximum should not be confused with the one due to soliton excitations
observed in some one-dimensional ferromagnets for $T\ll J$, which would cause a completely different magnetic
field dependence of $1/T_1$ \cite{Ferraro}. This low-frequency dynamic should rather be ascribed to the
progressive freezing of the molecular spins, as detected by means of AC susceptibility in the pristine
$\beta$-MnPc \cite{Yasu}. It should be remarked that a slowing down of the frequency of the spin fluctuations to
the MHz range at $T>\Theta$ can occur only if the magnetic anisotropy barrier $E_A=DS_z^2\geq \Theta$ and
overcomes the thermal energy $k_BT$, a situation which is similar to the one of single molecule magnetic
clusters\cite{Mn12} and which was observed also in $\alpha$-FePc \cite{Evangelisti}.

Now, if one assumes that the decay of the correlation function involving the transverse components of the
hyperfine field $<h_+(t)h_-(0)>= <\Delta h_{\perp}^2> exp(-t/\tau_c)$, with $<\Delta h_{\perp}^2>\simeq 2A^2
S(S+1)/3$, from Eq.3 one derives
\begin{equation}
\frac{1}{T_1}= \frac{\gamma^2}{2} <\Delta h_{\perp}^2> \frac{2\tau_c}{1+ \omega_N^2\tau_c^2} .
\end{equation}
Taking into account that the molecular anisotropy yields an effective energy barrier $E_A$, which in principle can
be slightly temperature dependent owing to the T-dependence of the in-chain correlation length, one can write
$\tau_c=\tau_o exp(E_A/T)$ ($E_A$ in Kelvin). If we try to fit $^1$H $1/T_1^s$ data by taking this expression for
$\tau_c$ in Eq.7 we obtain a poor fit. In fact, the broad maximum in $1/T_1^s$ indicates a distribution of
correlation times. For simplicity we consider a distribution of relaxation times associated with a rectangular
distribution of activation energies from $<E_A>-\Delta$ to $<E_A>+\Delta$. Then the average relaxation time turns
out \cite{Tedo}
\begin{eqnarray}
\frac{1}{T_1}= \frac{\gamma^2}{2} \frac{<\Delta h_{\perp}^2> T}{\omega_N\Delta}\biggl[
atan\biggl(\omega_N\tau_o e^{\frac{<E_A>+\Delta}{T}}\biggr) - \nonumber \\
-\, atan\biggl(\omega_N\tau_o e^{\frac{<E_A>-\Delta}{T}}\biggr)\biggr] .
\end{eqnarray}
A good fit of the data for different magnetic fields is obtained for $\tau_o= (1\pm 0.4)\times 10^{-10}$ s and
$<E_A>\simeq \Delta\simeq 25$ K $>\Theta$, as expected for a freezing of MnPc spins. It should be remarked that
this activation energy is quite different from the one (about 90 K) derived from the analysis of AC susceptibility
data. Since $1/T_1$ probes the $q$-integrated spin susceptibility \cite{RepProg}, this observation would indicate
a more rapid softening of the $q=0$ modes with respect to the other Brillouin zone modes.

Now we turn to the discussion of $^7$Li $1/T_1$. If one considers that the order of magnitude of the hyperfine
coupling of $^7$Li, derived from the line broadening (Fig. 10), is the same of $^1$H, one would expect also a peak
in $^7$Li $1/T_1$ around 10 K. Remarkably, $^7$Li $1/T_1$ does not show any peak, decreases upon cooling and is
much smaller than $1/T_1^s$ (see Figs. 6 and 8). On the other hand, this temperature dependence is close to the
one observed for $1/T_1^l$ the slowly relaxing component of $^1$H magnetization (Fig.8).

The absence of a low-temperature peak in $^7$Li $1/T_1$ cannot be due to a filtering of the spin excitations at
the critical wave-vector, as Li is not in a symmetric position with respect to Mn ions \cite{RepProg}, moreover
$1/T_1$ starts decreasing at $T\gg J$. Therefore, it is likely that in \lpc\, two different microscopic
configurations are present: Li-rich chain segments and Li depleted segments. The latter ones are characterized by
a spin dynamics similar to the one of pristine $\beta$-MnPc which is evidenced by $1/T_1^s$, while the former
regions show a quite different dynamics evidenced by $^7$Li $1/T_1$ and $^1$H $1/T_1^l$. The coexistence of two
different types of chains is consistent with the analysis of the magnetization data. It is worth to mention that
from high-resolution X-ray diffraction there is no evidence of a macroscopic segregation in Li-rich and
Li-depleted phases \cite{Yasu}, pointing out that the two configurations should coexist at the microscopic level.

At high temperature, around 180 K, $^7$Li $1/T_1$ shows a broad maximum (Fig. 9). One could argue that this
maximum is analogous to the one observed in $1/T_1^s$ at low temperature and that it could arise from a spin
freezing. However, at variance with what is found at $T\simeq 10$ K, no evidence of a spin freezing around $200$ K
is present in the magnetization data (see Fig.2). Another mechanism that could give rise to a maximum in $^7$Li
$1/T_1$ is Li diffusion, with a hopping frequency reaching the MHz range around 200 K. However, if this was the
case $^7$Li NMR linewidth should be much narrower and should not follow the same temperature dependence of the
macroscopic susceptibility (Fig. 10). Moreover, if Li was diffusing one should not observe a stretched exponential
recovery of nuclear magnetization with a temperature-independent exponent $\beta$ (Fig. 4). Therefore, $^7$Li
relaxation should originate from a different mechanism.

\begin{figure}[h!]
\vspace{6cm} \includegraphics{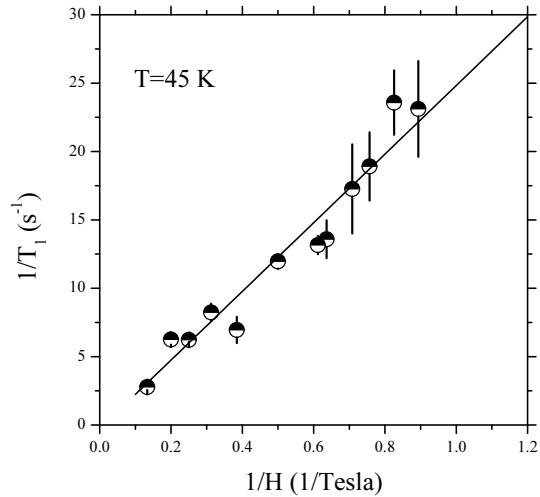} \caption{$^7$Li nuclear
spin-lattice relaxation rate, at $T=45$ K, reported as a function of $1/H$. The solid line shows the behaviour
expected from Eq.8 for $<E_g>=\Delta_g=410$ K . }\label{T1LivsH}
\end{figure}

In Li-rich chain segments the electrons should start filling the $2e_g(\pi)$ orbitals, which overlap with the
orbitals of adjacent molecules to form a one-dimensional band \cite{electro,Tosatti}. Nevertheless, the electron
delocalization can be prevented both by a strong Coulomb repulsion, as well as by the disorder associated with a
non-uniform Li distribution. Accordingly, an effective gap between localized and itinerant states develops.
Moreover, one should consider that the interaction of the electrons with localized spins can yield to
spin-polarons formation. In this scenario one can ascribe to the electron a phenomenological \cite{CuOLi} hopping
time $\tau_e=\tau_e^0 exp(E_g/T)$, with $\tau_e^0$ renormalized with respect to its bare value $\tau_e^0\simeq
\hbar/W$ ($W$ the bandwidth) owing to spin-polaron formation and other effects \cite{Zuppi}, while $E_g$ is an
effective gap between localized and delocalized states. The disorder, already evidenced by the stretched
exponential recovery, causes a distribution of $E_g$'s. If one assumes a rectangular distribution of $E_g$'s one
obtains again an expression for $1/T_1$ as in Eq.8. It is noticed that for a width of the distribution of the
order of its average value (i.e. $<E_g>\simeq \Delta_g$), at low temperature $1/T_1\propto T$. Although this
temperature dependence is the same one expected for a metal, here it has quite a different origin. Moreover, while
in a one-dimensional metal $1/T_1$ should be either frequency independent or increase as $1/\sqrt{H}$ \cite{Soda},
from Eq.8 $1/T_1$ should increase as $1/\omega_N\propto 1/H$, at low temperature. We find that the temperature and
magnetic field dependence of $1/T_1$ can be accounted for Eq.8 with $<E_g>\simeq \Delta_g\simeq 410$ K and a
$\tau_e^0\simeq 3\times 10^{-10}$ sec (see Figs. 8 and 11). Although it is not easy to give a definite statement
in favour of this relaxation mechanism on the basis of our data alone, we point out that this phenomenological
model for the electron diffusion along the chain can satisfactorily explain all $^7$Li NMR data. A homogeneous Li
distribution or, in other terms, less disorder would possibly allow to observe a metallic behaviour in Li$_x$MnPc.

\section{CONCLUSIONS}

From NMR  measurements we have addressed the study of the modifications of the microscopic properties of
$\beta$-MnPc upon doping with alkali-ions. We have observed that in \lpc\, two different dynamics coexist at the
microscopic level. A first one arising from the freezing of MnPc spins, which is similar to the one observed in
the undoped compound and is associated with Li depleted regions. A second one, characteristic of Li-rich regions,
associated with the hopping of the electrons along the chain, whose delocalization is hindered by the disorder,
the Coulomb and spin interaction. The temperature dependence of the characteristic correlation time for each one
of these two processes was derived and compared to the one obtained from macroscopic techniques. These
phthalocyanines could possibly become metallic for Li contents x=2 and x=4, where Li$_x$MnPc show a more
homogeneous distribution of the Li ions and less disorder is present.

\section*{Acknowledgements}

This work was supported by PRIN2004 National Project "Strongly Correlated Electron Systems with Competing
Interactions " and by Fondazione CARIPLO 2005 Scientific Research funds.



\end{document}